\RequirePackage{mathptmx}
\documentclass[a4paper,10pt,twoside,final,twocolumn]{svjour3}
\usepackage{graphicx}
\usepackage{booktabs}
\usepackage{amssymb,bm,mathrsfs,bbm,amscd}
\usepackage[tbtags]{amsmath}
\usepackage{newtxmath}
\usepackage[titletoc,title]{appendix}
\usepackage{lineno,hyperref}

\newif\ifpapercjk
\papercjkfalse

\newif\ifpaperthanks
\paperthankstrue

\newif\ifpaperpublic
\paperpublictrue

\ifpapercjk
\usepackage{CJK}
\fi

\begin{document}
\ifpapercjk
\begin{CJK*}{UTF8}{gbsn}
\fi

\title{Muon Tracking with the fastest light in the JUNO Central Detector
}

\author{
    Kun Zhang \and Miao He \and Weidong Li \and Jilei Xu
}

\institute{
  Kun Zhang \and Miao He \and Weidong Li \and Jilei Xu  \at
  Institute of High Energy Physics, Chinese Academy of Sciences, Beijing 100049, China  \\
  \email{hem@ihep.ac.cn} Miao He \\
  \and
  Kun Zhang \and Weidong Li \at
  University of Chinese Academy of sciences, Beijing 100049, China
}

\date{Received: date / Accepted: date}
\maketitle

\begin{abstract}
  The Jiangmen Underground Neutrino Observatory (JUNO) is a multi-purpose neutrino experiment designed to measure the neutrino mass hierarchy using a central detector (CD), which contains 20 kton liquid scintillator (LS) surrounded by about 18,000 photomultiplier tubes (PMTs), located 700~m underground. The rate of cosmic muons reaching the JUNO detector is about 3~Hz and the muon induced neutrons and isotopes are major backgrounds for the neutrino detection. Reconstruction of the muon trajectory in the detector is crucial for the study and rejection of those backgrounds. This paper will introduce the muon tracking algorithm in the JUNO CD, with a least squares method of PMTs' first hit time (FHT). Correction of the FHT for each PMT was found to be important to reduce the reconstruction bias. The spatial resolution and angular resolution are better than 3~cm and 0.4~degree, respectively, and the tracking efficiency is greater than 90\% up to 16~m far from the detector center.
\end{abstract}


\section{Introduction\label{sec:introduction}}

    The Jiangmen Underground Neutrino Observatory~\cite{Djurcic:2015vqa,An:2015jdp} is a multiple purpose neutrino experiment to determine neutrino mass hierarchy and precisely measure oscillation parameters, using reactor antineutrinos from Yangjiang and Taishan nuclear power plants.
Fig.~\ref{fig:detector_scheme} shows the schematic view of the JUNO detector. Twenty kiloton LS is contained in a spherical vessel with the radius of 17.7~m as the central detector (CD). The light emitted by the LS is watched by about 18,000 20-inch PMTs installed in the water pool, with the photocathode at a radius of 19.5~m, with more than 75\% optical coverage. There is a top tracker made of plastic scintillator bars, on top of the water pool.

\begin{figure}
  \centering
  \includegraphics[width=8cm]{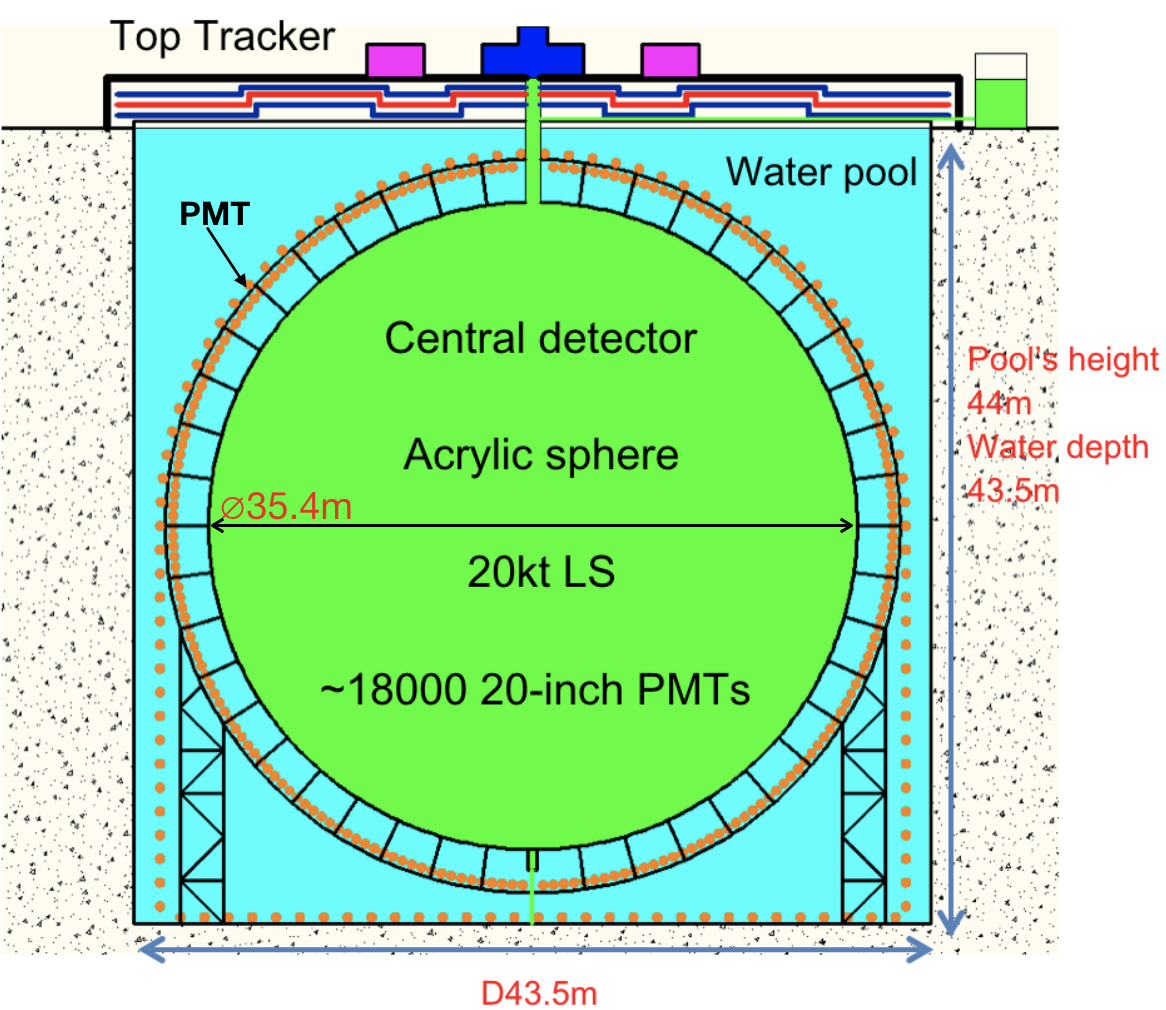}
  \caption{ Schematic view of the JUNO detector}
\label{fig:detector_scheme}
\end{figure}

In the JUNO detector, cosmogenic radioactive isotopes -- especially $^9$Li and $^8$He -- and fast neutrons are serious correlated background sources to reactor antineutrinos, which can be efficiently rejected by sufficient time veto after the tagged muons. For example, the time window to veto $^9$Li and $^8$He is required to be no less than 1.2~s, according to their lifetimes. However, according to the Monte Calor simulation, the rate of muon reaching CD is about 3~Hz with the mean energy of about 215~GeV. If the full LS volume is vetoed, there is almost no live time in the detector. Since the vertex of the cosmogenic isotopic is correlated with the primary muon in both time and space, the most effective way is to only veto a cylindrical volume along the muon trajectory instead of the full detector. This approach requests precise tracking of the muon. The  top tracker above CD can also track muons but it has small coverage and can only measure 25\% of the total cosmic muons, therefore, muon reconstruction in the CD is necessary for the entire 4$\pi$ solid angle.

This paper introduces the reconstruction of muons in the JUNO CD with the time signal of PMTs. The algorithm is described in detail in Sec.~\ref{sec:algorithm}, including the parameterization of a track, the prediction and correction of the first hit time (FHT), and the minimization with the least squares method. The reconstruction performance is shown in Sec.~\ref{sec:performance} and
the conclusion is in Sec.~\ref{sec:conclusion}.

\section{Algorithm\label{sec:algorithm}}

\subsection{Least Square Method\label{subsec:lsq}}

The straight track of a muon in the detector can be described by seven independent parameters: $x_{inj}$, $y_{inj}$, $z_{inj}$, $t_{inj}$, $\theta_p$, $\phi_p$, and $l_{trk}$. $x_{inj}$, $y_{inj}$, $z_{inj}$ and $t_{inj}$ are the position and time of the muon injecting into the LS with the center of the LS ball as the origin of coordinates. $\theta_p$ and $\phi_p$ are the direction of the muon track in spherical coordinates, and $l_{trk}$ is the length of muon trajectory in LS. Muons that are generated outside the detector and stopped inside are very rare in JUNO and are not considered in this paper. For muons going through the detector, the injection and outgoing points can be fixed at the surface of the LS sphere thus $x_{inj}$, $y_{inj}$, $z_{inj}$ and $l_{trk}$ can be replaced with another two parameters $\theta_{inj}$ and $\phi_{inj}$.

Then the FHT of PMT at position $\vec{R_i}$ can be predicted by the tracking parameters with a proper optical model:
\begin{equation}
  \centering
  T_i^{pre}=f\left(\vec{R_i};\theta_{inj},\phi_{inj},t_{inj},\theta_p,\phi_p\right)
\end{equation}

Details about the model will be discussed in Sec.~\ref{subsec:fht}. Then with the predicted FHT and the observed value, the $\chi^2$ can be built as:
\begin{equation}
  \centering
  \chi^2=\sum_i{\left(\dfrac{T_i^{pre}-T_i^{obs}}{\sigma_i}\right)^2},
\end{equation}

where $T^{obs}_i$ is the observed (from data or MC) value of FHT and $\sigma_i$ represents the error of FHT for the $i^{th}$ PMT.

The reconstructed track parameters are obtained by minimizing the $\chi^2$ function. Minuit2 in root package is used to minimize $\chi^2$, and initial values of the parameters are required. To get them, the position of the PMT which has the earliest FHT is regarded as the injecting point and its FHT as the injecting time, and the charge center of all PMTs (calculated by Eq.~\ref{equ:charge_center}) is calculated as the muon trajectory center in LS. With this information, the initial values of all the parameters can be inferred.

\begin{equation}
\centering
\vec{R}=\frac{\sum_i q_i\vec{R_i}}{\sum_i q_i}
\label{equ:charge_center}
\end{equation}

where $q_i$ and $\vec{R_i}$ are the charge and position of the $i^{th}$ PMT seperately.

\subsection{Fastest light model\label{subsec:fht}}
As shown in Fig.~\ref{fig:muonrec_schematic}, when a muon travels through the central detector, energy is deposited in the LS and scintillation lights are emitted isotropically along the muon track. There are also Cherenkov lights, however, most of them are absorbed by the LS and re-emitted as scintillation lights. As a result, only less than 1\% of lights detected by PMTs carries the directional information, which is extremely difficult to separate from scintillation lights. Therefore, all lights are treated as isotropic in our reconstruction.

After emission, the optical photon travel in the LS and water, and there is a certain probability for it to reach a PMT\@. For a specified PMT, among all the photons hit on it, the earliest one is defined as the fastest light of this PMT, and the time of the fastest light hit on the PMT is defined as the first hit time (FHT).

\begin{figure}
  \centering
  \includegraphics[width=5cm]{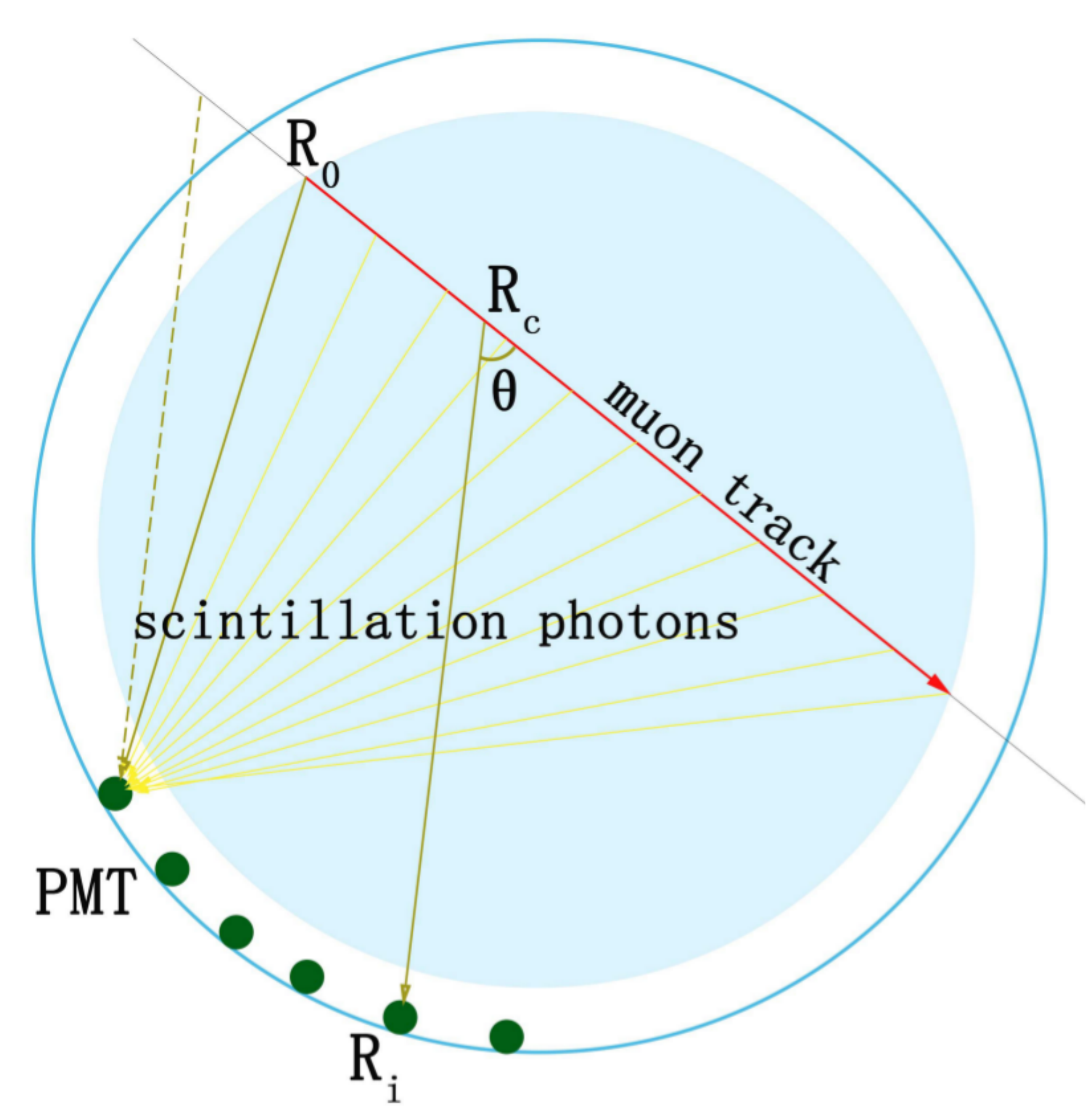}
  \caption{\label{fig:muonrec_schematic} Schematic view of fastest light. $\vec{R_0}$ is the injecting position of the track into the LS, $\vec{R_i}$ is the position of the $i^{th}$ PMT, and $\vec{R_c}$ is the position where the fastest light emitted for the PMT $\vec{R_i}$ }
\end{figure}

Given the time of muon injection $t_{inj}$, the time of a photon arriving at a certain PMT can be calculated as~\cite{Thomas:2011}:
\begin{equation}\label{eqn:hittime}
  \centering
t=t_{inj}+t_{\rm muon}+t_{\rm photon},
\end{equation}

where $t_{\rm muon}$ is the muon propagation time before the light emission, and $t_{\rm photon}$ is the time of flight of this photon. Defining the refractive index of LS as $n_{LS}$, the muon velocity as $c_{\mu}$, and the light velocity as $c$, it is straight forward to get
\begin{equation}
  \centering
t_{\rm muon}=\dfrac{l}{c_{\mu}},
\end{equation}

and
\begin{equation}
  \centering
t_{\rm photon}=\dfrac{\left|\vec{R_i}-\left(\vec{R_0}+l\hat{V}\right)\right|}{c/n_{LS}}.
\end{equation}

Here $l$ means the distance of the muon travelling before it emits that photon, and $\hat{V}$ is the unit vector of the muon track direction. Eq.~\ref{eqn:hittime} can be rewritten as
\begin{equation}
  \centering
  t=t_{inj}+\dfrac{l}{c_{\mu}}+\dfrac{\left|\vec{R_i}-\left(\vec{R_0}+l\hat{V}\right)\right|}{c/n_{LS}},
\end{equation}

and the minimum of $t$ can be obtained from the partial derivative with respect to $l$:
\begin{equation}
  \centering
  \label{eqn:emittingpoint}
  \begin{aligned}
    \dfrac{\partial{t}}{\partial{l}} & = \dfrac{1}{c_{\mu}}+\dfrac{n_{LS}}{c}\cdot\dfrac{\partial{\left|\vec{R_i}-\left(\vec{R_0}+l\hat{V}\right)\right|}}{\partial{l}}\\
  & = \dfrac{1}{c_{\mu}}-\dfrac{n_{LS}}{c}\cdot\cos\theta,
\end{aligned}
\end{equation}

where $\vec{R_{ci}}=\vec{R_i}-(\vec{R_0}+l\hat{V})$ means the vector from the emitting point to the PMT and $\hat{R_{ci}}$ is its unit vector, and $\theta$ is the angle between $\hat{R_{ci}}$ and $\hat{V}$. The calculation related with the vector's derivative in Eq.~\ref{eqn:emittingpoint} can refer to Appendix A. Considering for ultra-relativistic muons, $c_{\mu}\approx c$, to minimize $t$, there is:
\begin{equation}
  \centering
  \cos\theta=\dfrac{1}{n_{LS}}
\end{equation}

This means that for each PMT, the position where the fastest light emitted can be determined by the angle $\theta$,
which is the same as the emission angle of the Cherenkov light.
There is an exception that the calculated point $R_c$ is out of LS. In this case, the fastest light should come from the injecting point of muon into the LS $R_0$.

\subsection{Residual of the first hit time \label{subsec:fht-bias}}
\begin{figure}
  \centering
    \includegraphics[width=7cm]{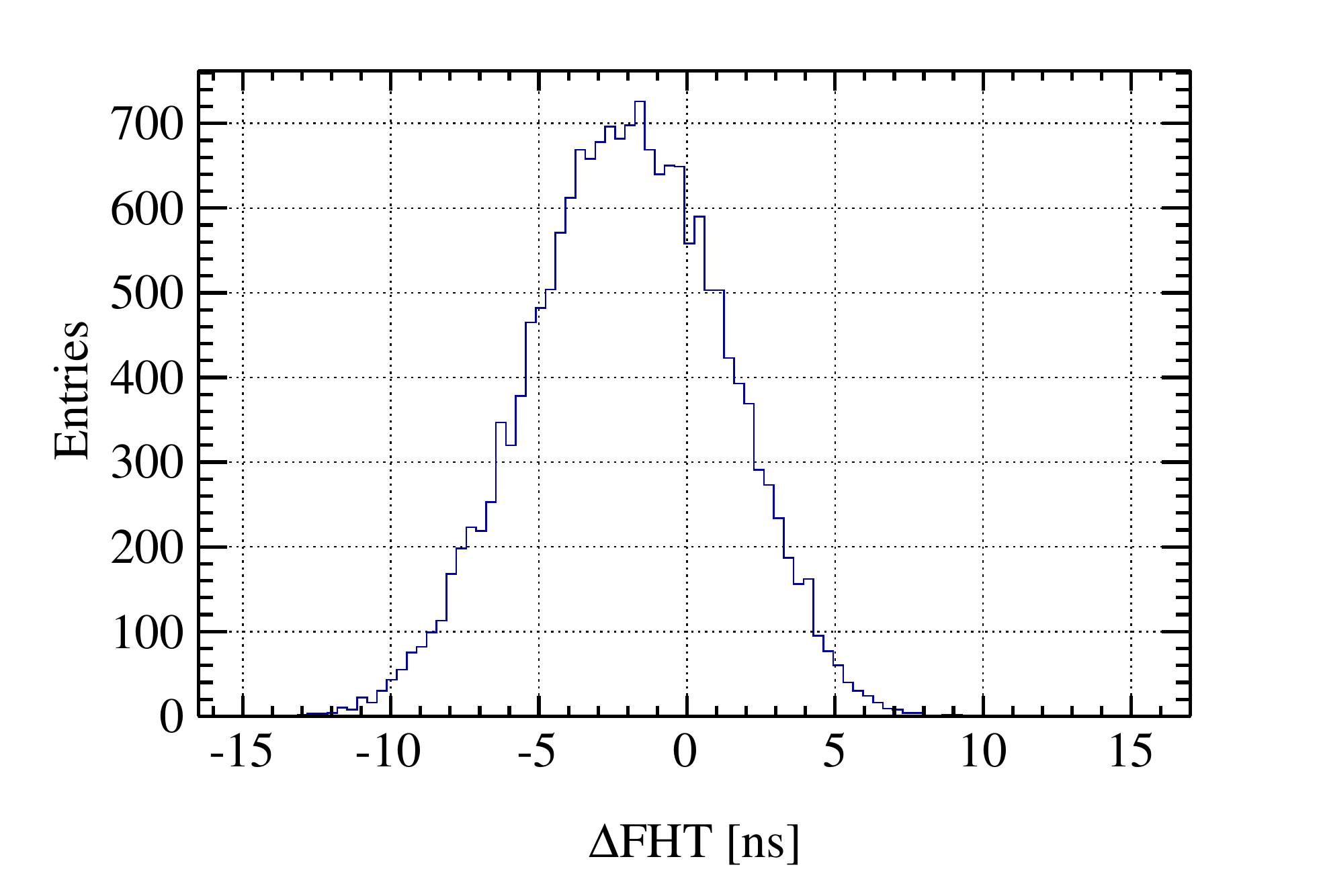}
  \caption{\label{fig:dt1evt}Residuals of FHTs of a MC muon event, with the TTS of PMT set as 3~ns. $\Delta{\rm FHT}=T^{obs}-T^{pre}$}
\end{figure}

Monte Calor simulation was done based on Geant4 with the JUNO detector geometry. A muon with 200~GeV kinetic energy was simulated going through the detector. The simulation software is a part of the JUNO offline software, and all the geometry parameters and optical properties used in the detector simulation are from JUNO Yellow Book\cite{An:2015jdp}.

The first hit time of each PMT obtained from MC truth was subtracted by the predicted time of the fastest light under the optical model in Sec.~\ref{subsec:fht} (using the track parameters in MC truth), and the distribution $\Delta$FHT is shown in Fig.~\ref{fig:dt1evt}.

There is not only fluctuation of $\Delta$FHT, but also an overall shift of the central value. There are a few reasons as below:

\begin{itemize}
  \item The scintillating light is not emitted instantaneously. Instead, the decay time follows a double-exponential distribution, with the time constants of 4.93~ns and 20.6~ns for the fast and slow components respectively.
  \item Because of different refractive index between LS and water, there should be reflection and refraction at the boundary of the LS ball, which are ignored in the optical model. In particular, when the light emission point is at the edge of the LS sphere, there is a region in which all PMTs can not see the light because of the total reflection. This effect is not included in the optical model either.
  \item There is an intrinsic transient time spread (TTS) of photoelectrons in the PMT. In this study, $\sigma_{\rm TTS}$ was assumed to be 3~ns, i.e. a 3~ns Gaussian smearing was added to the hit time of each photoelectron in MC.
  \item Scintillating lights are not infinite, and their emitting points are discrete along the track after all, which means that the fastest light for a given PMT does not have to come from the predicted position. Assuming the dE/dx of muon is 0.2~MeV/mm, and the average number of photoeletrons collected by all PMTs correlated with every MeV energy deposit in the CD center is about 1,200, considering the total number of PMT (20 inch) is about 18,000, along the track about $1200\times0.2/18000\approx0.013$ p.e./mm ($\rm p.e.\equiv photoelectron$) can hit on each PMT on average. This means that on average the muon emits a photoeletron hit for one PMT when flying every $1/0.013\approx 77.0mm$ long, which can be considered as the mean uncertainty of the position at which the fastest light is emitted, and the corresponding uncertainty of FHT is about $77mm/c_\mu\approx0.26$ ns, which is acceptable. From another view, if we want the uncertainty of FHT smaller than 1~ns, we need the muon emitting 1~p.e. when flying every $c_\mu\times1ns$ for each PMT. And for a trajectory through the CD center, with the track length of about 35,000~mm, this means that $35000mm/(c_\mu\times1ns)\approx120 ~\rm p.e.$ are needed for each PMT. This is a strict PMT selecting condition when reconstructing and 100~p.e. cut condition is used in the performance study in Sec.~\ref{sec:performance}.
  \item For a cosmic muon, there are multiple photoelectrons in each PMT. Assuming all hits are from the same point source and the probability density function (PDF) of the single hit time, after the time-of-flight subtract, is $f(x)$, then the PDF of FHT in case of $n$ photoelectrons is: $F(t,n)=nf(t)\left(\int_{t}^{+\infty}{dxf(x)}\right)^{n-1}$. However, for a muon, photons may be from any point along the track, thus the PDF of FHT can not be analytically expressed.
\end{itemize}

As a result, an additional correction to the predicted FHT is needed, as well as its fluctuation.

\subsection{Correction to the first hit time\label{subsec:bias-correction}}

In this study, MC muon samples were used to correct the first hit time of each PMT, taking into account the full detector geometry, with all optical parameters taken from Ref.~\cite{An:2015jdp}. In reality, there is a top tracker which can detect a small fraction of cosmic muons going through the LS ball with very good tracking resolution, which can be used as a calibration source to tune MC.

According to the analysis above, the FHT biases strongly depend on the number of Photoelectrons collected on each PMT. In addition, the relative position between the PMT and the track also have significant impacts on the FHT distribution. We defined some parameters as below to identify each PMT.

\begin{figure}
  \centering
    \includegraphics[width=3.7cm]{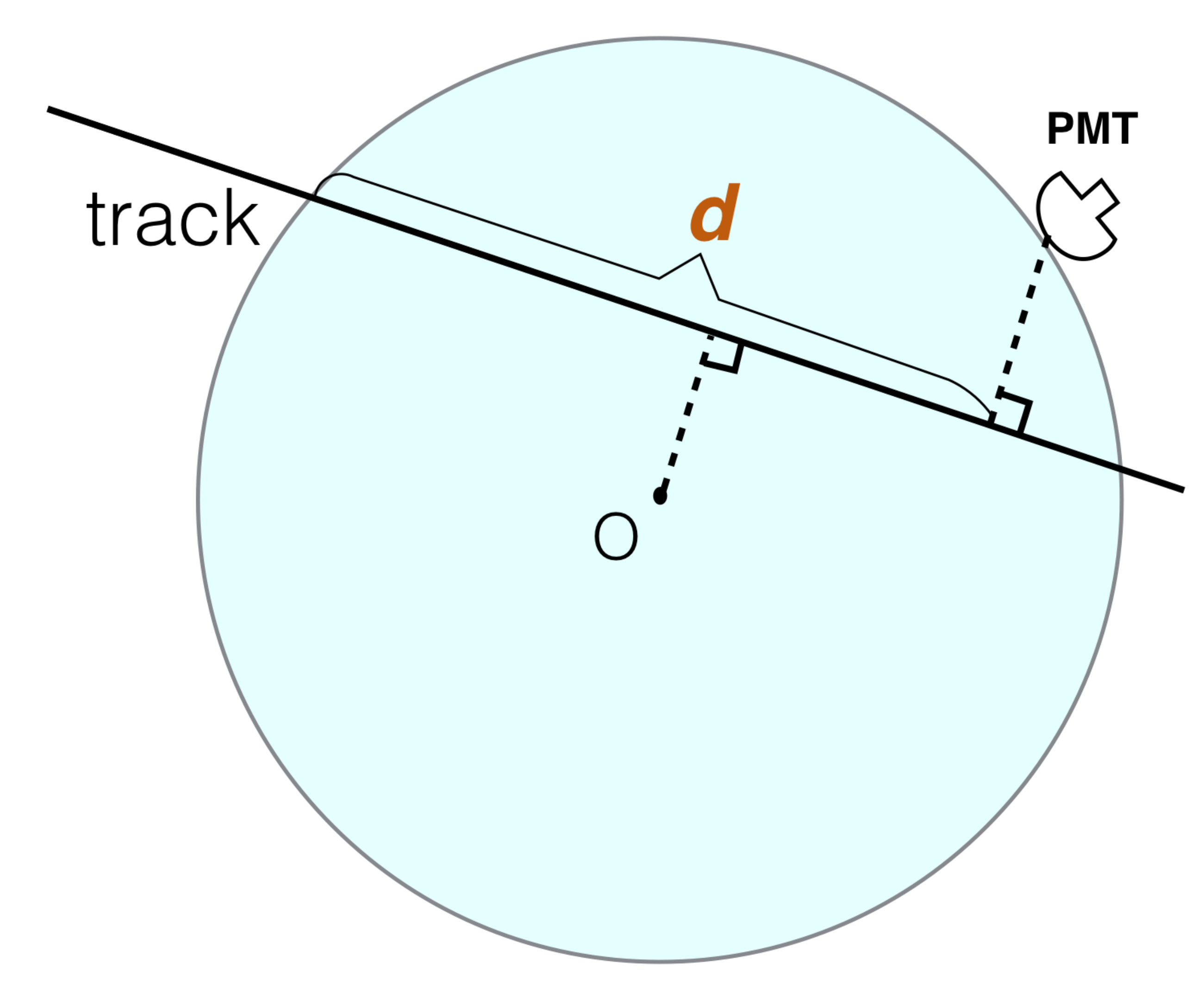}
    \includegraphics[width=4cm]{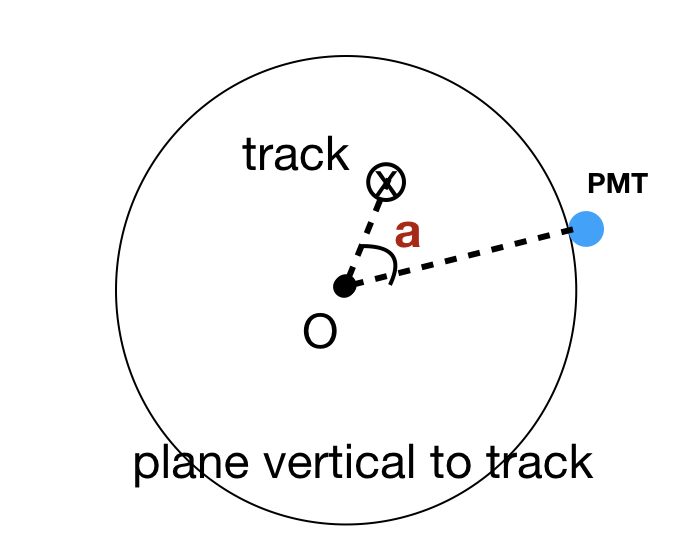}
  \caption{\label{fig:definition-d} Definition of correcting factors $d$ and $a$ }
\end{figure}

\begin{itemize}
    \item[-] $\bm{d}$: as shown in the left panel of Fig.~\ref{fig:definition-d}, the distance between the injecting point to the projection of the PMT falling at the track, which ranges from -19.5~m (when the track is at the edge of the LS ball, half of PMTs have negative $d$) to 37.2~m (when the track goes through the center of CD).
    \item[-] $\bm{a}$: the azimuth angle of each PMT, in the plane perpendicular to the track, which ranges from $0$ to $\pi$ (the area $-\pi\sim0$ can be combined into that of $0\sim\pi$ for the symmetry).
    \item[-] $\bm{D(ist)}$: the distance between the track and the CD center.
    \item[-] $\bm{q}$: the number of Photoelectrons.
\end{itemize}

Apparently, with a given $d$, $q$ and $a$ are correlated, thus we tried two different combinations $a$-$d$ and $a$-$d$ to study the correction of FHT. The correcting factors $\Delta T^{\rm pre}$, defined as the mean of $\Delta FHT=T^{\rm obs}-T^{\rm pre}$, as a function of $q$-$d$ and $a$-$d$, are shown in Fig.~\ref{fig:dt-factors} (a) and (b), respectively. From Fig.~\ref{fig:dt-factors}(a), we can see that the correcting factor becomes small when $q$ increases as expected, because of the first hit selection. On the other hand, given a fixed $q$, the correcting factor changes with $d$, because the detected photons have different emission points along the track. In Fig.~\ref{fig:dt-factors}(b), the left top region corresponds to the total reflection area where the first hit can not come from the predicted emission point thus have a latency, while the top middle region is furthest from the track so the PMTs have least $q$.

With these samples we can also obtain the standard deviation of FHT in every bin, which can be treated as the effective error of FHT $\sigma_i^{\rm eff}$ in the $\chi^2$ function Eq.~\ref{equ:corr-chi2}.

\begin{equation}
  \centering
    \label{equ:corr-chi2}
  \chi^2=\sum_i{\left(\dfrac{T_i^{pre}+\Delta T_i^{pre}-T_i^{obs}}{\sigma_i^{\rm eff}}\right)^2}
\end{equation}

Since both $d$ and $a$ rely on the tracking parameters, in principle, they have to be re-calculated in every iteration during the minimization process. However, the variation of the correcting factors and the effective error of the FHT makes the minimization unstable and difficult to converge. Therefore, to solve this technical problem in practice, we only run the correction and minimization twice and we found the results are acceptable.

\begin{figure}
  \centering
  \includegraphics[width=7cm]{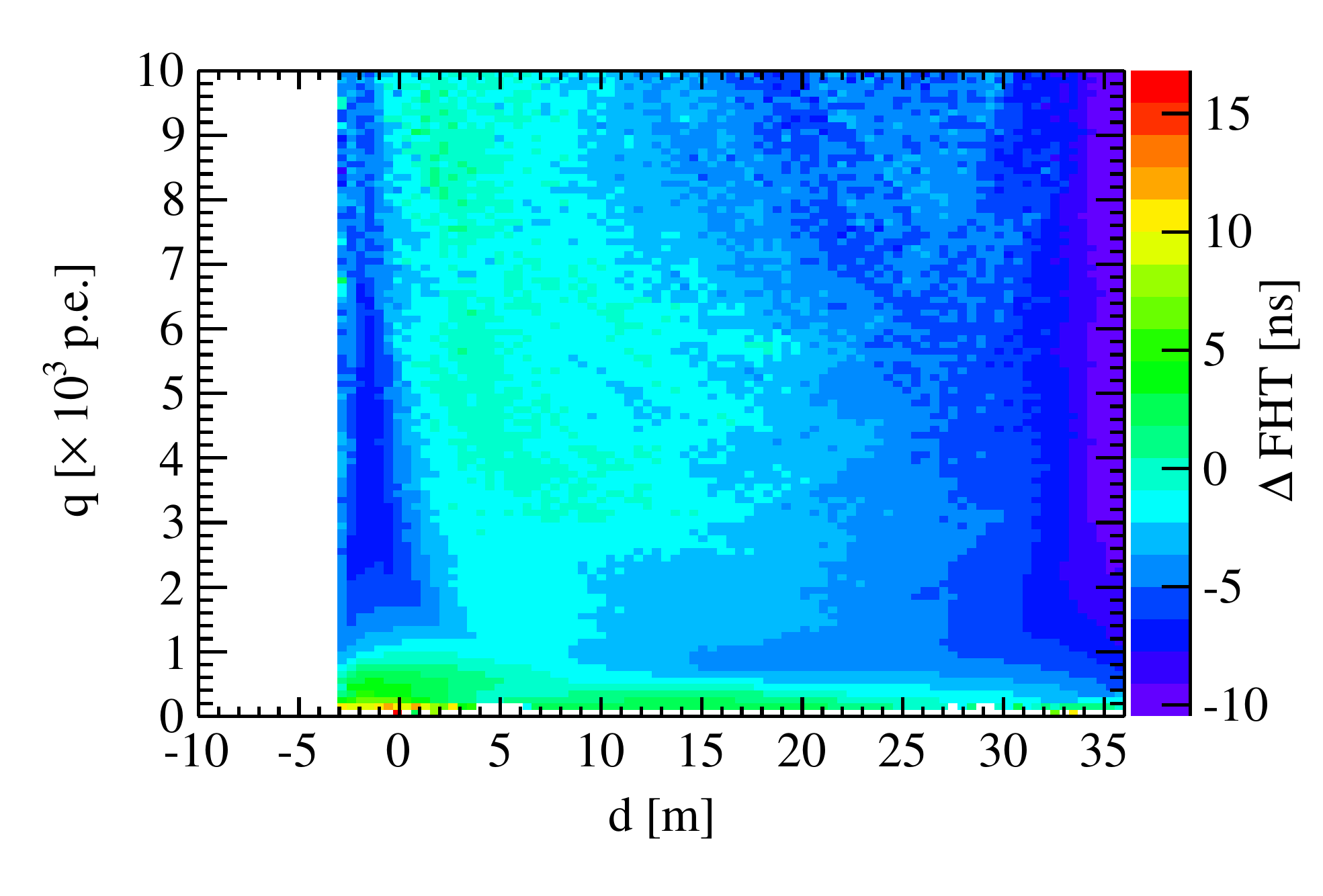} \\
  (a) $\Delta$FHT vs $q$ and $d$  \\
  \includegraphics[width=7cm]{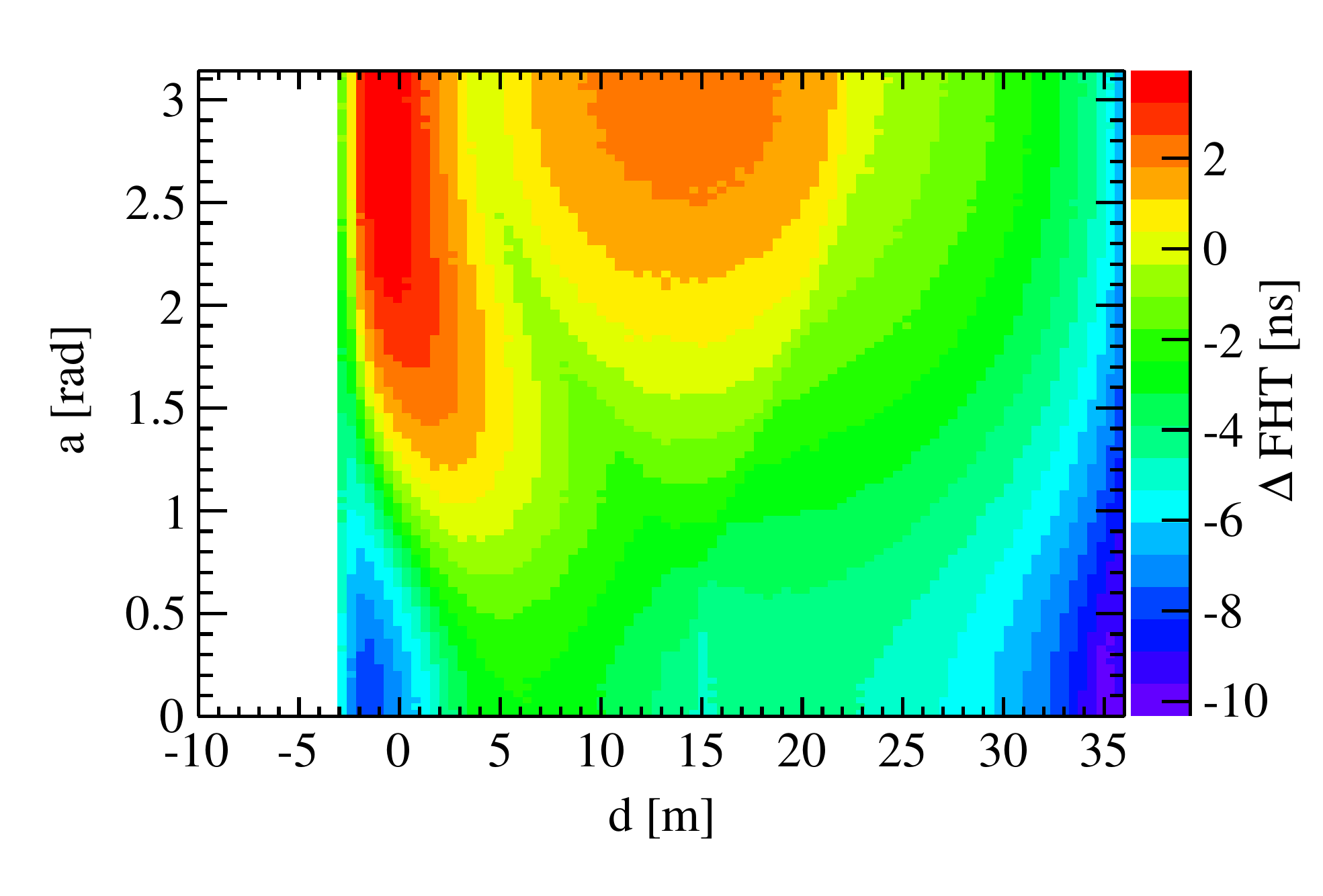} \\
  (b) $\Delta$FHT vs $a$ and $d$
  \caption{\label{fig:dt-factors} $\Delta$FHT vs the correcting factors. Samples are MC muon tracks 5.5~m away from the CD center with initial momentum 200~GeV and time resolution 3~ns}
\end{figure}

\section{Performance study with MC data\label{sec:performance}}

\subsection{Comparison of different FHT correction methods\label{subsec:perform-corr}}

The FHT correction mainly aims to solve the reconstruction biases. In this section we will evaluate the biases by $\alpha$ which is the angle between the reconstructed track and the MC truth one, and $\Delta D$ which means the error of reconstructed $D(ist)$.

Fig.~\ref{fig:rec-compare} compares the different performances with different FHT correcting methods.  We can see the mean $\alpha$ and $D(ist)$ without any correction can reach several degrees and dozens of centimeters respectively. And the only $q$ based correcting method can not improve the performance too much.

Correcting methods ``corr1d XXX'' means correcting the FHT by the parameters XXX one by one with relevant one-dimensional correcting curves, while the ``corr2d XXX'' methods are correction by all the two correcting-parameters at the same time with a corresponding two-dimensional correcting map.

It is easy to see that the four kinds of correction methods by 2 parameters perform better. And the best one is "corr2d a-d" method, with average $\alpha$ angle smaller than $0.4^{\circ}$ and average $\Delta D(ist)$ smaller than 1~cm.

\begin{figure}
  \centering
  \includegraphics[width=7cm]{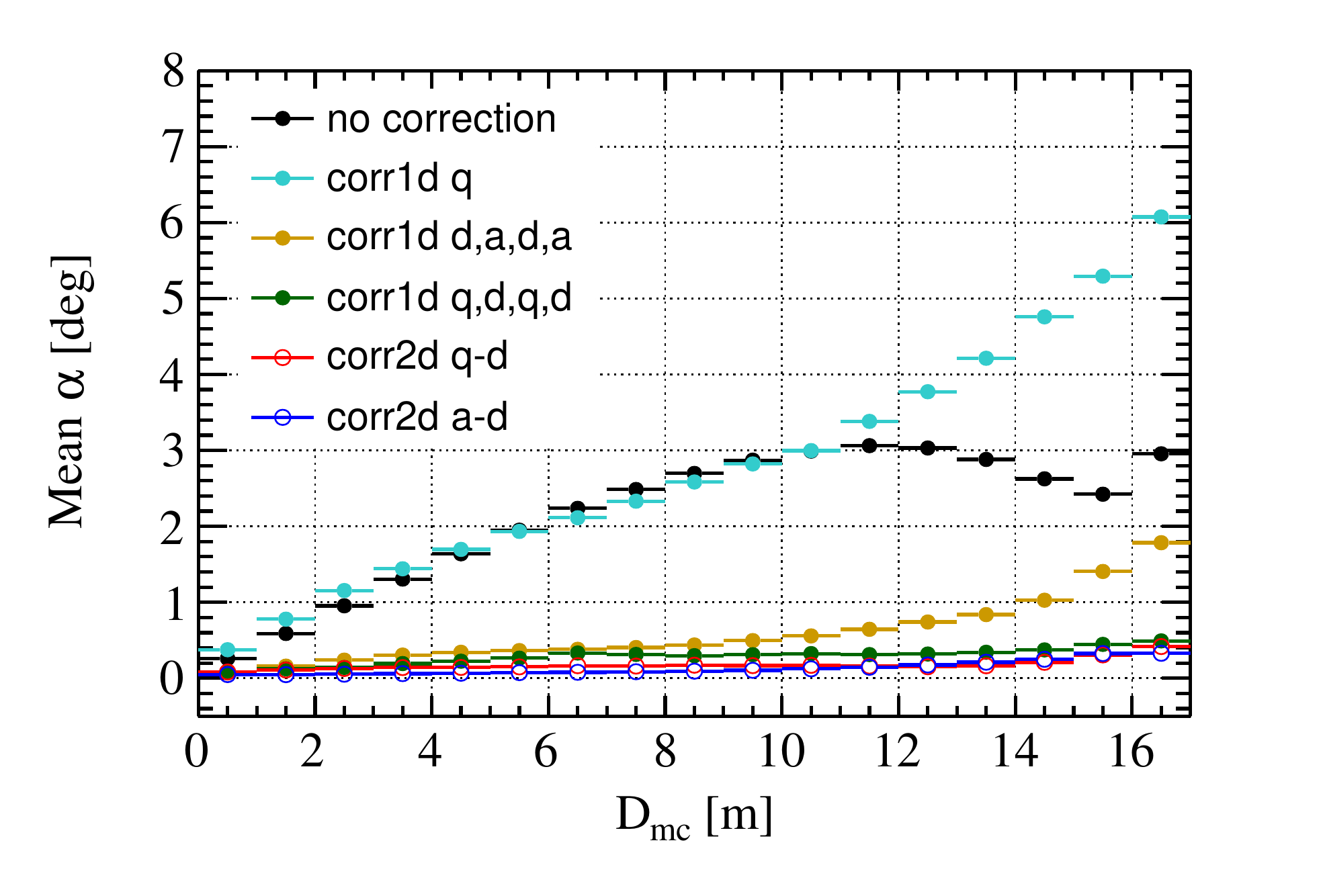} \\
  (a) Mean value of $\alpha$ vs true $D(ist)$ \\
  \includegraphics[width=7cm]{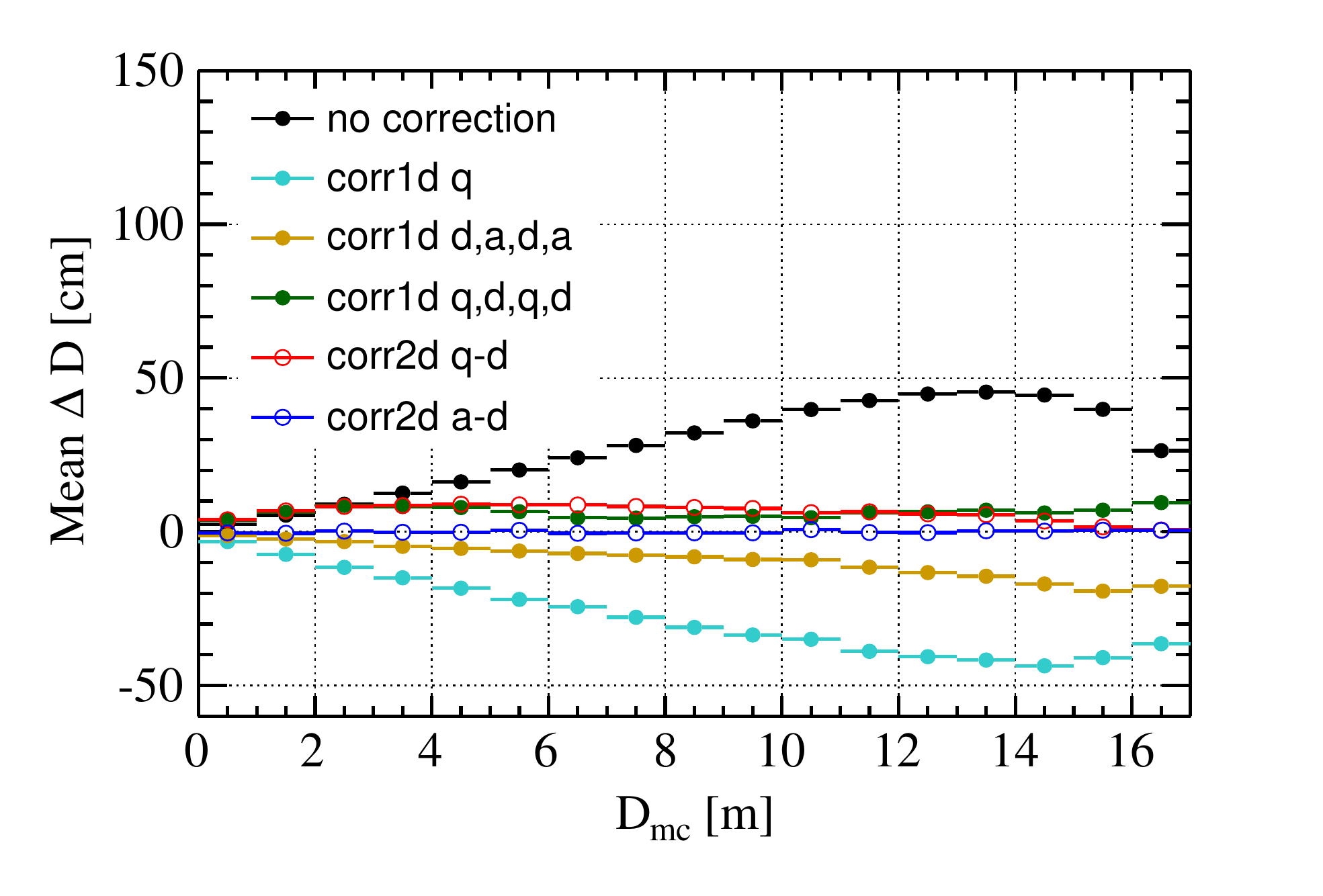} \\
  (b) Mean value of biases of reconstructed $D(ist)$ vs true $D(ist)$\\
  \caption{\label{fig:rec-compare} Performances of reconstruction with various FHT correcting methods. Samples are MC data with PMTs' TTS of 3ns}
\end{figure}

\subsection{Reconstruction resolution and efficiency\label{subsec:resolution_and_eff}}

In this section, the spatial and angular resolution and the tracking efficiency are shown using the correction method "corr2d a-d".

Fig.~\ref{fig:res-and-eff} shows the reconstruction resolution and the efficiency. The ``injecting x,y,z'' means the position of the track injecitng into the LS. And the resolutions are figured out by a gaussian fitting on data in every bins. From Fig.~\ref{fig:res-and-eff}~(a) we can see the spatial resolution is better than 3~cm and the angular resolution is better than 0.4~degree, up to 16~m far from the detector center. Muons going through the edge are very difficult to be reconstructed because the injection and outgoing points are too close to be separated. In Fig.~\ref{fig:res-and-eff}~(b), successful reconstruction is defined with all the 5 track parameters are in the range of 5 times of their standard deviation. With the distance less than about 14 meters of the track to the CD center, the tracking efficiency can be greater than 90\%, and with tracks close to the edge of the LS, the efficiency will decrease about 10\%.

\begin{figure}
  \centering
    \includegraphics[width=7cm]{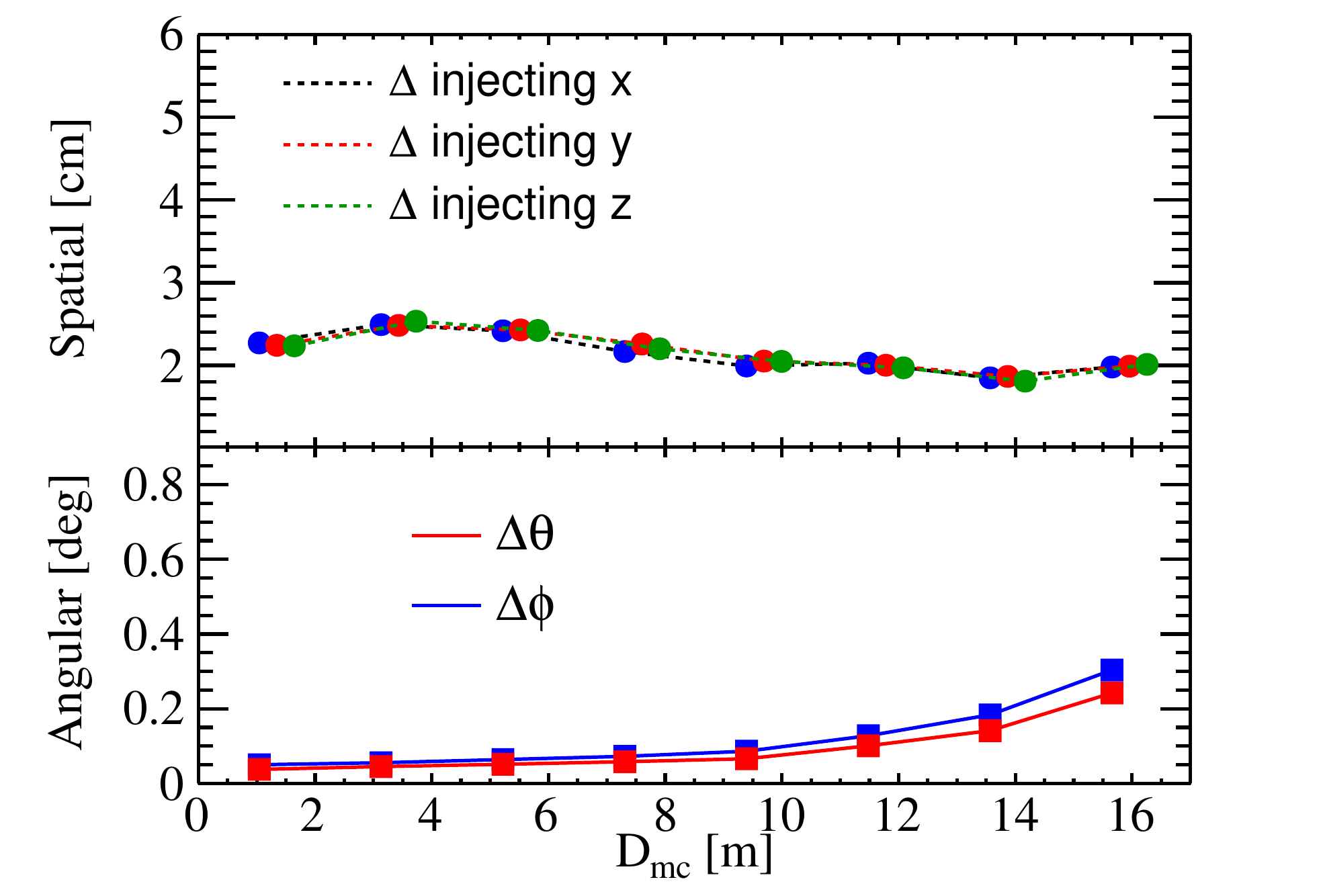}\\
    (a) Reconstruction spatial and angular resolution \\
    \includegraphics[width=7cm]{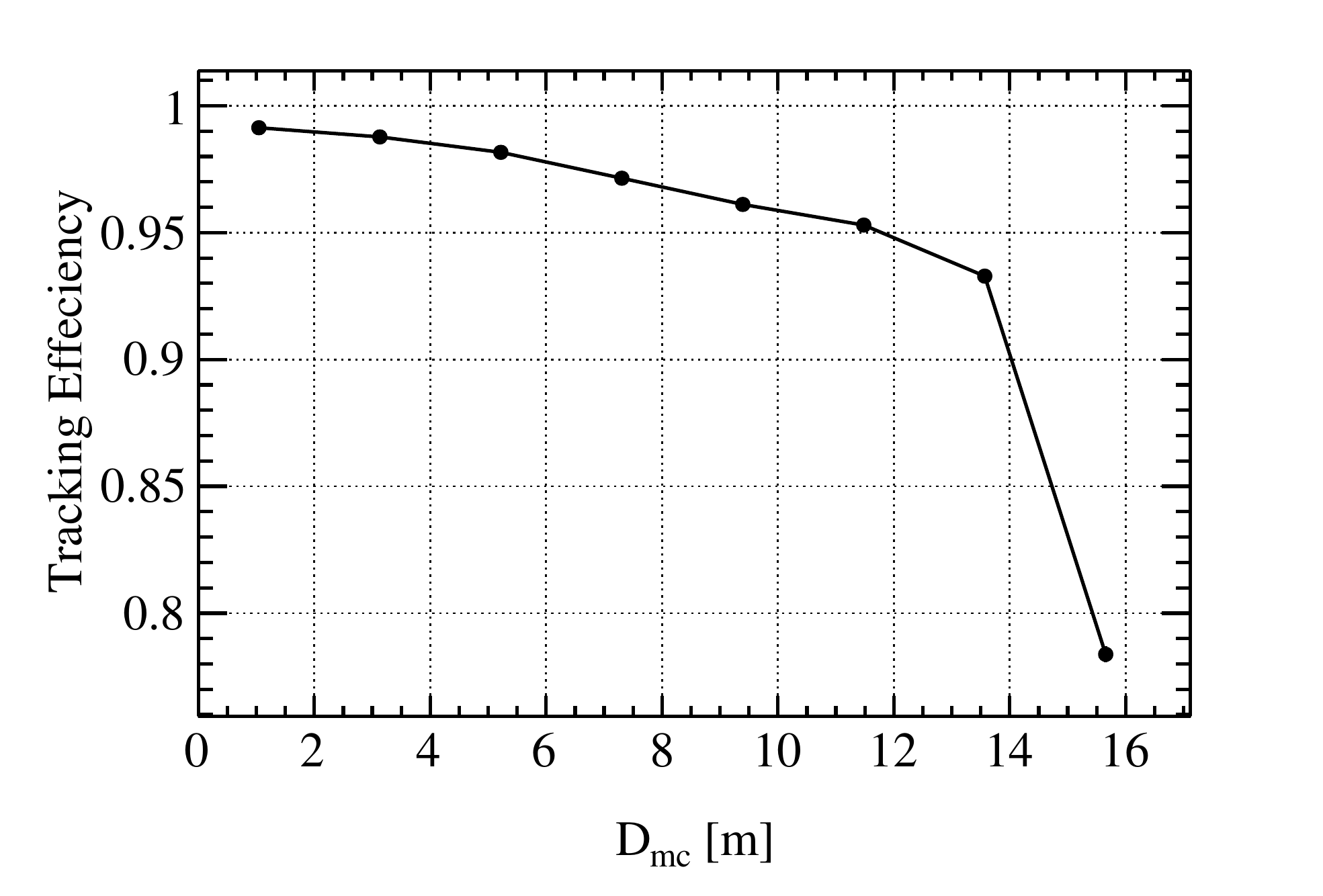}\\
    (b) Reconstruction efficiency
    \caption{\label{fig:res-and-eff} Reconstruction resolution and efficiency with correction method ``corr2d a-d''. In order to do comparison easily, the x axes are shifted a little for $\Delta$injecting-y and $\Delta$injecting-z in the top figure. }
\end{figure}

\section{Conclusion\label{sec:conclusion}}

We have developed a reconstruction algorithm of muon in JUNO CD based on the least square method and the fastest light model, but which has biases on account of the inaccurate prediction of the FHT. Considering that the top-tracker in JUNO above the CD can track some tracks downwards from top, we can use them to obtain the FHT-correction data vesus related observed correcting-factors, and then with which correct the predicted FHT in the reconstruction. With this algorithm, we can effectively reconstruct the muon track in the CD almost without biases and the spatial and angular resolution can be better than 3~cm and 0.5~degree respectively. And in most cases the tracking efficiency can be better than 90\%. More studies on the fastest light model will be done to get better prediction of the FHT and thus to improve the reconstruction performances.
\ifpaperthanks
\begin{acknowledgements}
    This work is supported by National Natural Science Foundation of China (Grant No. 11575226, 11605222), Joint Large Scale Scientific Facility Funds of NSFC and CAS (Grant No. U1532258) and the Strategic Priority Research Program of the Chinese Academy of Sciences (Grant No. XDA10010900)
\end{acknowledgements}
\fi

\subsection*{Appendix A\label{sec:appendix}}

\noindent{\bf Derivation of $\bm{\partial{\left|\vec{A}-l\hat{V}\right|}/\partial{l}}$  }

let $\vec{R}=\vec{A}-l\hat{V}$, then
\begin{equation*}
	\begin{aligned}
		\dfrac{\partial{\left|\vec{A}-l\hat{V}\right|}}{\partial{l}} & = \dfrac{\partial\left|\vec{R}\right|}{\partial{l}} = \dfrac{\partial{\left(\left|\vec{R}\right|^2\right)^{\frac{1}{2}}}}{\partial{l}} \\
			& =\frac{1}{2}\cdot\left(\left|\vec{R}\right|^2\right)^{-\frac{1}{2}}\cdot\dfrac{\partial{\left|\vec{R}\right|^2}}{\partial{l}} \\
		  & =\dfrac{1}{2}\cdot\dfrac{1}{\left|\vec{R}\right|}\cdot\dfrac{\partial{\vec{R}^2}}{\partial{l}} \\
	%\end{aligned}
	%\begin{aligned}
			& =\dfrac{1}{2}\cdot\dfrac{1}{\left|\vec{R}\right|}\cdot\dfrac{\partial{\left(\vec{A}^2-2l\vec{A}\cdot\hat{V}+l^2\hat{V}^2\right)}}{\partial{l}} \\
		  & =\dfrac{1}{2}\cdot\dfrac{1}{\left|\vec{R}\right|}\cdot-2\left(\vec{A}-l\hat{V}\right)\cdot\hat{V} \\
			& =-\dfrac{\vec{R}}{\left|\vec{R}\right|}\cdot\hat{V} =-\hat{R}\cdot\hat{V} \\
			& =-\cos<\hat{R},\hat{V}>
	\end{aligned}
\end{equation*}
where $\hat{R}$ means the unit vector of $\vec{R}$ and $<\hat{R},\hat{V}>$ means the angle between $\hat{R}$ and $\hat{V}$. For the partial derivative in Eq.~\ref{eqn:emittingpoint},
\begin{equation*}
	\vec{A}=\vec{R_i}-\vec{R_0}, \\
	\vec{R}=\vec{R_{ci}}=\left(\vec{R_i}-\vec{R_0}\right)-l\hat{V}
\end{equation*}
Therefore,
\begin{equation*}
	\dfrac{\partial{\left|\vec{R_i}-(\vec{R_0}+l\hat{V})\right|}}{\partial{l}} = -\hat{R_{ci}}\cdot\hat{V} = -\cos\theta
\end{equation*}

\ifpapercjk
\end{CJK*}
\fi

\begin{thebibliography}{90}
\newcommand{\etal}{et al}
\bibitem{Djurcic:2015vqa}
    Z. Djurcic \etal{} (JUNO Collaboration), arXiv:1508.07166
\bibitem{An:2015jdp}
    F.P. An \etal{} (JUNO Collaboration), arXiv:1507.05613,
    J. Phys. G: Nucl. Part. Phys. 43 (2016) 030401.
\bibitem{Thomas:2011}
    T.M. O'Donnell,
    Precision Measurement of Neutrino Oscillation Parameters with KamLAND,
    Ph.D. thesis,
    University of California,
    2011
\end{thebibliography}
\end{document}